 \documentclass[final,5p,times,twocolumn]{elsarticle}
\usepackage{amsmath}

\usepackage{amssymb}
\def\bC {\mathbb{C}}
\newcommand{\SsZk}{{S^7/{\mathbb{Z}}_m}}
\newcommand{\dd}{\textrm{d}}
\newcommand{\Hypergeometric}[4]{\,{}_2F_1\left(#1,#2;#3;#4\right)}

\newcommand{\NN}{\mathcal{N}}

\newcommand{\Cset}{\bC}

\journal{Physics Letters B}

\begin{document}

\begin{frontmatter}



\title{On the Regularization of Extremal Three-point Functions Involving Giant Gravitons}

\author{Charlotte Kristjansen$^a$ , Stefano Mori$^a$, and Donovan Young$^b$ }
\address[label1]{Niels Bohr Institute, Copenhagen University,\\ Blegdamsvej 17, 2100 Copenhagen \O, Denmark}
\address[label2]{ Centre for Research in String Theory, School of Physics and Astronomy, \\ Queen Mary University of London
\\ Mile End Road, London E1 4NS, United Kingdom}


\begin{abstract}
In the $AdS_5/CFT_4$ set-up, extremal three-point functions involving
two giant 1/2 BPS gravitons and one point-like 1/2 BPS graviton, when
calculated using semi-classical string theory methods, match the
corresponding three-point functions obtained in the tree-level gauge
theory. The string theory computation relies on a certain
regularization procedure whose justification is based on the match
between gauge and string theory.  We revisit the regularization
procedure and reformulate it in a way which allows a generalization to the
ABJM set-up where three-point functions of 1/2 BPS operators are not
protected and where a match between tree-level gauge theory and
semi-classical string theory is hence not expected.

\end{abstract}


\end{frontmatter}



After the successful application of integrability techniques to the planar spectral problem of the $AdS_5/CFT_4$
set-up \cite{Beisert:2010jr}, the calculation of three-point functions in the same set-up has attracted renewed attention
with some recent highlights being the conjecture of an all loop formula for three-point functions of single trace operators
in certain sub-sectors of ${\cal N}=4$ SYM \cite{Basso:2015zoa} and  the formulation of certain integrability axioms for the cubic string theory vertex \cite{Bajnok:2015hla}.

We will be considering three-point functions which do not belong to the class of three-point functions considered in the above references. Our three-point functions involve giant gravitons which in the string theory language correspond
to higher dimensional D- or M-branes wrapping certain submanifolds of the string theory background and which in the
gauge theory picture are represented by specific linear combinations of multi-trace operators, namely Schur polynomials. 
Remaining in the gauge theory picture, our three-point functions will involve two Schur polynomials and one single
trace operator, all of  $1/2$ BPS type.  Furthermore, the three operators will be chosen such that $\Delta_1=\Delta_2+\Delta_3$, where the $\Delta$'s
are the conformal dimensions of the operators. Such three-point functions are denoted as extremal 
three-point functions and are known to require special care in the comparison between gauge and string theory~\cite{D'Hoker:1999ea}. On the gauge theory side the three-point functions of interest can be calculated using techniques from zero-dimensional
field theories~\cite{Bissi:2011dc}  (see also~\cite{Corley:2001zk})
 and on the string theory side they can be determined by  generalizing a method developed
for the calculation of heavy-heavy-light correlators~\cite{Zarembo:2010rr,Janik:2010gc,Costa:2010rz} from string states to 
membranes \cite{Bissi:2011dc}.

In the case of the $AdS_5\times S^5$ correspondence the $1/2$-BPS nature of the operators involved implies that 
the three-point function considered is protected and thus should take the same value whether calculated in string theory
or in gauge theory.  As pointed out in~\cite{Caputa:2012yj} the need for special treatment of extremal correlators in
string theory is relevant here and in~\cite{Lin:2012ey} a  regularization procedure for the string theory computation
which led to the desired match  between gauge and string theory was presented. 

The $AdS_4\times \Cset \mathrm{P}^3$ set-up~\cite{Aharony:2008ug} allows one to consider a similar correlator i.e.\ an extremal three-point function involving two 1/2 BPS giant gravitons in combination with one  1/2 BPS point-like graviton and the methods developed in~\cite{Bissi:2011dc} for the $AdS_5/CFT_4$ calculation can be generalized to this case as well~\cite{Hirano:2012vz}. One remaining subtle
point is the choice of regularization procedure in the string theory computation. In the $AdS_4\times \Cset \mathrm{P}^3$ correspondence
three-point functions of 1/2 BPS operators are not protected and hence in this
set-up we can not expect a match between gauge and string theory results.  In particular, this means that on one hand  
we can not justify our choice of regulator by a match between the gauge and string theory results but on the other hand 
a  computation of the correlator in the weakly coupled string theory  will provide us with a non-trivial prediction about the behaviour of the correlator in the dual strongly coupled field theory.  Below we will revisit the the regularization procedure
employed for the $AdS_5\times S^5$ computation and modify it in a way that allows us to generalize it to the 
$AdS_4\times \Cset \mathrm{P}^3$ case. Subsequently, we carry out the string theory calculation of the extremal three-point function involving two giant 1/2 BPS gravitons and one point-like 1/2 BPS graviton in $AdS_4\times \Cset \mathrm{P}^3$. 

\section{Giant graviton correlators in $AdS_5\times S^5$ revisited}

Giant gravitons in $AdS_5\times S^5$ are D3 branes which wrap an $S^3$ which constitutes a subset of either $AdS_5$ or  $S^5$~\cite{McGreevy:2000cw,Grisaru:2000zn,Hashimoto:2000zp}.   We will consider the giants for which the wrapped sphere $S^3\subset S^5$ and which spin along a circle in the $S^5$ while being located at the center of $AdS_5$. For such giants
the dual gauge theory operators are Schur polynomials  built on completely anti-symmetric Young diagrams and containing
a single complex scalar field that we will denote as $Z$~\cite{Balasubramanian:2001nh,Corley:2001zk}. 
The  D3-brane action is (in units where the AdS radius has been set to 1)
\begin{equation}
S_{D3} = -\frac{N}{2\pi^2}\int d^4\sigma \left( \sqrt{-g} - P[C_4] \right), 
\end{equation}
where $g_{ab} = \partial_a X^M \partial_b X^N G_{MN}$, with  $a,b=0,\ldots,3$ being
the worldvolume coordinates and $X^M$ the embedding
coordinates.
Working in global $AdS$ coordinates 
\begin{align}\label{global}
ds^2  = &-\cosh^2 \rho\, dt^2 + d\rho^2 + \sinh^2\rho\, d\widetilde{\Omega}_3^2 \nonumber \\
&+ d\theta^2 + \sin^2\theta\, d\phi^2 + \cos^2\theta \,d\Omega_3^2,
\end{align}
the four-form potential $C_4$ can be written as~\cite{Grisaru:2000zn}
\begin{align}
C_{\phi \chi_1 \chi_2 \chi_3} =\, \cos^4\theta\, \text{Vol}(\Omega_3),
\end{align}
where the $\chi_i$ are the angles of the wrapped sphere, i.e.\
$d\Omega_3^2=d\chi_1^2+\sin^2\chi_1 d\chi_2^2+\cos^2\chi_1 d\chi_3^2$.
 Using the ansatz
\begin{equation} \label{ansatz}
\rho= 0,\quad \sigma^0 = t,\quad \phi = \phi(t),\quad \sigma^i = \chi_i,
\end{equation}
one can show that a D3-brane with angular momentum $k$ is stable when it sits at
\begin{equation}
\cos^2\theta=k/N,
\end{equation}
and spins at the speed of light, $\dot{\phi}=1$\cite{McGreevy:2000cw}.
In order to determine the three-point function of two giant gravitons and a point-like graviton (i.e.\ a chiral primary)  we 
should determine the variation of the Euclidean version of the $D3$ brane action in response to the insertion of the 
desired chiral primary at the boundary of $AdS$ and subsequently evaluate these fluctuations on the Wick rotated giant
graviton solution. As this procedure was described in detail in~\cite{Bissi:2011dc} we shall be brief here. Denoting the
spherical harmonic representing  the point-like graviton as $Y_{\Delta}$ (with $\Delta$ referring to its conformal dimension)
we can write the variation of the D3-brane action as~\cite{Bissi:2011dc}
\begin{eqnarray}
 \delta
S&=&\frac{N}{2\pi^2}\cos^2\theta\int d^{\,4}\sigma
\left(\frac{2}{\Delta+1}Y_\Delta\,
(\partial^2_t -\Delta^2)\, s^{\Delta} \right. \nonumber \\
&&\left.+ 4 \label{deltaS}
\left[\Delta \cos^2\theta-\sin\theta \cos\theta\,  \partial_{\theta}
\right]
s^\Delta Y_\Delta \right),
\end{eqnarray}
where $s^{\Delta}$ is the bulk to boundary propagator. As our spherical harmonic we choose
\begin{equation}\label{simple}
Y_\Delta = \frac{\sin^\Delta\theta \,e^{i\Delta\phi}}{2^{\Delta/2}},
\end{equation}
which corresponds to the single trace operator $\mbox{Tr}\,Z^{\Delta}$ in the gauge theory language.
With this choice for $Y_\Delta$ the first term in eqn.\ (\ref{deltaS}) is finite and gives the following contribution
to the three-point function structure constant
\begin{equation}
C^3_{finite}= -\sqrt{\Delta} \, \frac{k}{N} \left(1-\frac{k}{N}\right)^{\Delta/2},
\end{equation}
whereas the contribution coming from the term with square brackets takes the form of a divergent integral with a pre-factor
which tends to zero. In reference~\cite{Lin:2012ey} it was proposed to regularize the divergent integral by replacing the
simple spherical harmonic $Y_\Delta$ with the more involved one
\begin{equation}\label{extendedspherical}
Y_{\Delta+2l,\Delta}={\cal N}_{\Delta,l} \sin^{\Delta}\theta\, e^{i\,\Delta\, \phi}\,  {_2F_1}(-l,\Delta+l+2;\Delta+1;\sin^2\theta),
\end{equation}
where ${\cal N}_{\Delta,l}$ is a normalization factor, 
and to consider the limit $l\rightarrow 0$ where $Y_{\Delta+2l, \Delta}\rightarrow Y_{\Delta}$, and where the contribution
from the ill defined term of eqn.\ (\ref{deltaS})  becomes finite. This procedure leads to a match between the string and
gauge theory computation. The obtained match justifies the choice of the regularization procedure but does not suggest a general principle that one could build on when aiming at a generalization to  giant gravitons in $AdS_4\times  \Cset \mathrm{P}^3$. One 
property which characterizes the spherical harmonic~(\ref{extendedspherical}) is that it extends the simple one without
making use of additional coordinates on $S^5$. However, this property is somewhat deceptive and is not the
correct clue to an extension to the $AdS_4\times  \Cset \mathrm{P}^3$ set-up. 

Here we shall formulate the regularization procedure in a slightly different manner which will allow us to generalize it to
the latter set-up. For that purpose we make use of the fact
 that spherical harmonics on $S^5$ are in one-to-one correspondence with
symmetric traceless $S\!O(6)$ tensors. In particular, (leaving out normalization factors) the spherical harmonic~(\ref{simple}), which translates into $\mbox{Tr}\, Z^\Delta$ in the field theory, corresponds to the tensor
\footnote{The chiral primary operators of ${\cal N}=4$ SYM can be written in the form  $C_I^{i_1i_2\ldots i_{\Delta}}\mbox{Tr}
(\Phi_{i_1}\Phi_{i_2}\ldots \Phi_{i_\Delta})$ where the $\Phi_i$'s can be any of the six real scalar fields and where
$C_I$ is a symmetric traceless tensor. We take the complex scalar field  $Z$ to be given by $Z=\Phi_1+i\Phi_2$.}
\begin{equation}
C_{\underbrace{1\ldots 1}_{\Delta-k}\underbrace{2\ldots 2}_k}=i^{\,k},
\end{equation}
where symmetrization is understood.  It is easy to show that adding more indices of type 1 and type 2 to the tensor (i.e.\ adding more fields of type
$\Phi_1$ and $\Phi_2$ to the operator) does not regularize the divergent integral. However, one can regularize
the integral by considering the following symmetric traceless tensor 
\begin{equation}\label{regulator}
C_{\underbrace{1\ldots 1}_{\Delta-k}\underbrace{2\ldots 2}_{k}\underbrace{3\ldots 3}_{2l-n}\underbrace{4\ldots 4}_{n}}=i^{\,k+n},
\end{equation}
where $n<2l$ and subsequently taking the limit $l\rightarrow 0$.  Obviously, the gauge theory operator resulting from
this tensor involves the complex field $Y=\Phi_3+i\Phi_4$ in addition to the complex field $Z$. It is easy to check that the
spherical harmonic~(\ref{extendedspherical}) corresponds to an operator involving all six scalar fields of ${\cal N}=4$ SYM
but it is not straightforward to express the corresponding $C$-tensor in a closed form. 
The tensor~(\ref{regulator}) translates into  the following spherical harmonic
\begin{equation}
Y_{\Delta \,l}={\cal N}_{\Delta\, l} \sin^{\Delta}(\theta) \, e^{i \Delta \phi} \cos^2(l\, \theta) \, \sin^2 (l\, \chi_1) \,
e^{2 i \,l\, \chi_2 },
\end{equation}
where ${\cal N}_{\Delta\, l}$ is another normalization constant.
Using this spherical harmonic instead of $Y_{\Delta}$ when evaluating the second line of~(\ref{deltaS}) and subsequently
taking the limit $l\rightarrow 0$ gives us the following result for the regularized contribution to the three-point function
\begin{equation}
C^3_{regularized}=\frac{1}{\sqrt{\Delta}}\, \left( 1+\frac{k}{N}\Delta\right) \left(1-\frac{k}{N}\right)^{\Delta/2},
\end{equation}
meaning that the total three-point function takes the form
\begin{equation}
C^3_{total}=\frac{1}{\sqrt{\Delta}} \left(1-\frac{k}{N}\right)^{\Delta/2},
\end{equation}
which precisely coincides with the gauge theory result obtained in~\cite{Bissi:2011dc}. One can also extend the $C$-tensor
in eqn.~(\ref{regulator}) to include indices of type $5$ and $6$ corresponding to the scalar fields $\Phi_5$ and
$\Phi_6$. This leads to the same result for the regularized contribution to the three-point function.

\section{Giant graviton correlators in $AdS_4\times  \Cset \mathrm{P}^3$ }
The giant graviton that we will consider is most easily described in the M-theoretic language where it corresponds to
an M5 brane which wraps two $S^5$'s  intersecting at an $S^3$, all inside $\SsZk$.
Furtermore, the M5-brane rotates 
along a circle orthogonal to the compactification circle. \footnote{For a detailed
explanation of the $AdS_4\times  \Cset \mathrm{P}^3$ set-up we refer to~\cite{Aharony:2008ug}.}
The general version of this
brane was constructed in~\cite{Giovannoni:2011pn} but maximal versions 
appeared already in ~\cite{Berenstein:2008dc,Murugan:2011zd}. We shall be considering the general case for which an
improved parametrization was found in~\cite{Hirano:2012vz} and we shall follow the notation of the latter reference.  The field
theory dual to this giant graviton is a Schur polynomial which involves two of the complex bi-fundamental scalar fields of ABJM theory and is built on the completely anti-symmetric Young tableau. For details we refer to~\cite{Hirano:2012vz,Giovannoni:2011pn}, see also~\cite{Dey:2011ea,Chakrabortty:2011fd}. The giant graviton is a 1/2 BPS object and we will be interested in calculating the three-point function involving
two such giants and a point-like graviton, likewise 1/2 BPS.  The point-like graviton translates in the field theory language 
to a single trace operator involving the same two complex bi-fundamental fields as the giant. Unlike what is the case in
the $AdS_5\times S^5$ set-up, three-point functions of 1/2 BPS operators are not protected within the $AdS_4\times \Cset \mathrm{P}^3$
correspondence and a match between tree
level gauge theory and semi-classical string theory results is not to
be expected. Thus the more reason there is
to be careful about the choice of regulator for the string theory computations. 

To describe the giant graviton it is convenient to use the  parametrization of $\SsZk$ introduced in~\cite{Hirano:2012vz}
\begin{align}
 Z_1&=r e^{\,\rho+i\theta} e^{i(\phi+\frac{\chi}{2})}, & Z_2&=r e^{\,-\rho+i\theta} e^{i(\phi-\frac{\chi}{2})}, \label{coord1} \\
  Z_3&=e^{\rho_3+i(\phi+\theta_3)}, & Z_4&=r_4 e^{i\phi},\label{coord2}
\end{align}
where \(r_4^2=1-e^{2\rho_3} -2r^2\cosh 2\rho\) and where the ranges of the coordinates are as follows
\begin{align}
 &0\leq r\leq \frac{1}{\sqrt{2}}, \hspace*{0.3cm}
  \rho_3\leq\frac{1}{2}\log(1-2r^2\cosh 2\rho),\\
 -&\rho_{max}\leq\rho\leq\rho_{max} \hspace*{0.3cm} \text{with }\hspace*{0.2cm} \cosh2\rho_{max}=\frac{1}{2r^2},\\
 &0\leq \theta,\,\theta_3,\,\chi,m\phi\,\leq 2\pi.
\end{align}
The $Z_i$ are then in one-to-one correspondence with the four complex bi-fundamental fields of ABJM theory. 
The giant graviton is described
by the equation
\begin{equation}
Z_1\overline{Z}_2 =\alpha^2 e^{i\chi(t)},
\end{equation}
where $t$ is the global $AdS$ time and $\alpha$ is a constant. Referring to the
parametrization introduced in~(\ref{coord1}) and~(\ref{coord2}),  the world volume coordinates of the brane are
$(t,\rho,\rho_3, \theta,\theta_3,\phi)$ where 
\begin{align}
-&\rho_M\leq \rho\leq \rho_M, \hspace{0.3cm} \cosh(2\rho_M)=\frac{1}{2\alpha^2}, 
\end{align}
and the other ranges are as before. The brane is maximal for $\alpha=0$ and minimal for 
$\alpha=\frac{1}{\sqrt{2}}$. The low energy effective action of the M5 giant is a sum of a DBI and a WZ term which
in the given parametrization can be written as follows~\cite{Hirano:2012vz} 
\begin{align}
 S_{\textrm{DBI}}&=-c\cdot\alpha^2\int\dd^6\sigma \, e^{2\rho_3}\sqrt{(\cosh 2\rho-2\alpha^2\omega^2)(\cosh 2\rho-2\alpha^2)},
 \nonumber\\
S_{\textrm{WZ}}&=c\cdot \alpha^2\int\dd^6\sigma \,e^{2\rho_3}(\cosh 2\rho -2\alpha^2)\omega,
\end{align}
where $c=R_\SsZk^6/\left((2\pi)^5 l_p^{\,6}\right)$ and $\omega=\dot{\chi}(t)$.
As shown numerically in~\cite{Giovannoni:2011pn} the Routhian of the system is minimized for $\omega=1$. Hence,
like its $AdS_5\times S^5$ cousin the present giant 
spins with the speed of light.
The corresponding angular momentum 
$k$ is related to its size in the following 
way~\cite{Hirano:2012vz}
\begin{equation}
 \frac{k}{N}=\sqrt{1-4\alpha^4}-4\alpha^4\log\frac{1+\sqrt{1-4\alpha^4}}{2\alpha^2}.
\end{equation}
To calculate the three-point function involving two giant gravitons and a point-like graviton (a chiral primary) we should again evaluate the variation of the Euclidean action in response to the insertion of the chiral primary at the boundary of $AdS$. Denoting as before the spherical harmonic representing the chiral primary as $Y_{\Delta}$ we can write the variation of
the action is
\begin{align} \label{deltaSDBI}
\delta S_{DBI}&=c\cdot\int\dd^6\sigma \,\frac{1}{2} \sqrt{g}\,\Biggl(2\Delta+  \nonumber\\
&\frac{\cosh 2\rho}{\cosh 2\rho-2\alpha^2\omega^2}
          \left[\frac{4}{\Delta+2}\partial_t^2-\frac{\Delta^2}{\Delta+2}\right]
\Biggr)Y_\Delta(\Omega)s^\Delta(X),\\
          \label{deltaSWZ}
\delta S_{WZ}&=-c\cdot \int\dd^6\sigma \omega 
          \left.\sqrt{g_\SsZk} g_\SsZk^{r\beta}\partial_\beta Y_\Delta(\Omega)\right|_{r=\alpha} s^\Delta(X).
\end{align}
As our spherical harmonic we would like to use
\begin{align}
 Y_{\Delta}=\left(r^{2} e^{i\chi}\right)^{\Delta/2}, \label{S7simple}
\end{align}
which corresponds to the single trace operator $\mbox{Tr}(Z_1\overline{Z}_2)^{\Delta/2}$ containing the same fields
as the giant graviton. With this choice of spherical harmonic the contribution to the three-point function from the term
containing square brackets is finite and yields
\begin{align}
{\cal C}^3_{finite}=\frac{1}{N} \left( \frac{\lambda}{2\pi^2}\right)^{1/4} (2k)\,  (2\alpha^2)^{\Delta} \sqrt{2\Delta+1},
\end{align}
where $\lambda=N/m$. The remaining part of the contribution from eqn.~(\ref{deltaSDBI}) and~(\ref{deltaSWZ}) is ill-defined.
In the same manner as it was the case for the $AdS_5\times S^5$ set-up the ill-defined part takes the form of a divergent integral times
a vanishing pre-factor. This again calls for a regularization of the contribution. With an eye to the regularization of 
reference~\cite{Lin:2012ey} one would be tempted to look for a regulator in the form of a solution of the Laplace equation
which would reduce to the simple one (\ref{S7simple}) when some of its quantum numbers were sent to zero and which would 
involve as few as possible additional coordinates compared to (\ref{S7simple}). However, the Laplace equation when expressed in the coordinates~(\ref{coord1})--(\ref{coord2}) can only in a few special cases be solved by separation of variables and the corresponding spherical harmonics turn out not to regularize the three-point function. 

On the other hand, the method based on using appropriate tensors as the starting point for the regularization, works neatly in the $AdS_4\times \SsZk$ case as well and, in addition, allows us to understand why certain spherical harmonics fail to
regularize the three-point function. The spherical harmonic (\ref{S7simple}) corresponds to the tensor \footnote{The chiral primary operators of ABJM theory can be written in the form
$(C_I)^{i_1\ldots i_{\Delta/2}}_{j_1\ldots j_{\Delta/2}}\mbox{Tr}(Z_{i_1}\overline{Z}^{j_1}\ldots Z_{i_{\Delta/2}}
\overline{Z}^{j_{\Delta/2}})$ where the $Z_i$'s can be any of the four complex fields of the theory and where $C_I$
is symmetric in upper and (independently) in lower indices and the trace taken over any pair of one upper and
one lower index vanishes.} 

\begin{equation}
 C_{\underbrace{2\ldots2}_{\Delta/2}}^{\overbrace{1\ldots1}^{\Delta/2}}=1.
\end{equation}
As in the $AdS_5\times S^5$ case one can regularize the ill defined contribution to the three-point function by dressing the
tensor above with  extra indices corresponding to fields not already present in the chiral primary~(\ref{S7simple}) and sending the number of extra indices
to zero at the end of the calculation. A choice for the extension of the tensor indices which mimics closely the one of the
$AdS\times S^5$ is
\begin{equation}\label{S7regulator}
 C_{\underbrace{2\ldots2}_{\Delta/2}\underbrace{3\ldots3}_{l-k}\underbrace{4\ldots4}_{k}}^{\overbrace{1\ldots1}^{\Delta/2}\overbrace{3\ldots3}^{l-k}\overbrace{4\ldots4}^{k}}=(-1)^{\,k},
 \end{equation}
where symmetrization is understood. This tensor corresponds to the spherical harmonic
\begin{equation}
 Y_{\Delta l}=\widetilde{\NN}_{\Delta l} \, r^{\,\Delta}\ e^{i\,\chi\Delta/2}\ e^{2l\rho}\ \Hypergeometric{-l}{-l}{1}{e^{2\rho_3}r_4^2}.
\end{equation}
Using this spherical harmonic instead of $Y_{\Delta}$ from eqn.\ (\ref{S7simple}) when evaluating the divergent part of (\ref{deltaSDBI}) 
and~(\ref{deltaSWZ}) and sending $l$ to zero at the end of the calculation leaves one with the following 
contribution to the three-point function
\begin{equation}
{\cal C}^3_{regularized}=-\frac{\Delta+2}{\Delta} \,\,  {\cal C}^3_{finite},
\end{equation}
so that the full three-point function in the $AdS_4\times \Cset \mathrm{P}^3$ case becomes \footnote{We have left out
a sign here since three-point functions are anyway only determined up to a phase factor.}
\begin{align}
{\cal C}^3_{total}=\frac{1}{N} \left( \frac{\lambda}{2\pi^2}\right)^{1/4} (2k)\,\,  (2\alpha^2)^{\Delta} \,\,
\left(\frac{ 2 \sqrt{2\Delta+1}}{\Delta}\right),
\end{align}
A natural question to ask is what would happen if we chose an even more general tensor as a regulator such as
the tensor
\begin{equation}
  C_{\underbrace{2\ldots2}_{\Delta/2}\underbrace{1\ldots1}_{k_1} \underbrace{2\ldots2}_{k_2}\underbrace{3\ldots3}_{k_3}\underbrace{4\ldots4}_{k_4}}%
  ^{\overbrace{1\ldots1}^{\Delta/2}\overbrace{1\ldots1}^{k_1}\overbrace{2\ldots2}^{k_2}\overbrace{3\ldots3}^{k_3}\overbrace{4\ldots4}^{k_4}}
  =(-1)^{\,k_2+k_4},
\end{equation}
where \(\sum_{i=1}^4 k_i=l\). For such tensors it is possible to show that the regularization procedure leads to negative
powers of $\alpha$ in the three-point function unless $k_1=k_2$. Negative powers of $\alpha$ in the three-point function
implies a divergence of the  three-point function in the maximal limit and the correlator is thus not fully regularized.  Enforcing
$k_1=k_2$ it is not possible to construct a traceless symmetric tensor unless $k_1=k_2=0$ which brings us back to
the previous case~(\ref{S7regulator}). 

The regularization procedure suggested in~\cite{Lin:2012ey}  for extremal three-point functions involving  two
giant gravitons and a chiral primary in $AdS_5\times S^5$  could be justified by its implication of an expected match between three-point functions of 1/2 BPS objects in  string and gauge theory. As already mentioned, in the $AdS_4\times \Cset \mathrm{P}^3$ set-up three-point functions of 1/2 BPS operators are not protected and thus one has to rely on the analogy with the $AdS_5\times S^5$ case when choosing a  regularization procedure.  By considering the procedure suggested in \cite{Lin:2012ey} from a slightly different angle we have been able to construct  a generalization for the $AdS_4\times \Cset \mathrm{P}^3$ case.  The procedure consists of dressing the tensor describing the chiral primary involved in the three-point function with indices of a type not already present,  performing the calculation of the correlator and sending the quantum numbers counting the extra indices to zero. 
We notice that the outcome of this procedure is  very similar to that of the $AdS_5\times S^5$ case in that the extra contribution to the three point function coming from  the regularized integral takes the form of a pre-factor times the finite part of the three-point function.

\section*{Acknowledgments}
C.\ Kristjansen was supported by FNU through grant number DFF -- 1323
-- 00082. 
S. Mori was supported by Fondazione ``Angelo Della Riccia''.
D. Young was supported by the Science and Technology
Facilities Council Consolidated Grant ST/L000415/1 {\it ``String
  theory, gauge theory \& duality"} and by NORDITA. He also thanks the
organisers of the ESF and STFC supported workshop ``Permutations and
Gauge String duality'' held at Queen Mary University of London where
some of this work was presented and discussed. We thank S.\ Hirano and
A.\ Martirosian for participating in the initial phases of this
project. We furthermore thank A.\ Bissi, H. Lin, A. Prinsloo, and
K. Zoubos for discussions.




\end{document}